\documentclass[5p]{elsarticle}

\usepackage{amssymb,amsmath}
\usepackage[dvips]{color}
\newcommand{\be}{\begin{equation}}
\newcommand{\ee}{\end{equation}}
\newcommand{\bea}{\begin{eqnarray}}
\newcommand{\eea}{\end{eqnarray}}
\newcommand{\beas}{\begin{eqnarray*}}
\newcommand{\eeas}{\end{eqnarray*}}

\newcommand{\nn}{\nonumber}
\newcommand{\slsh}[1]{{\not \! #1}}

\journal{Physics Letters B}

\begin{document}

\begin{frontmatter}

\title{Inverse magnetic catalysis from the properties of the QCD coupling in a magnetic field}

\author{Alejandro Ayala$^{1,2}$, C. A. Dominguez$^2$, L. A. Hern\'andez$^2$, M. Loewe$^{2,3,4}$, R. Zamora$^{5,6}$}

\address{$^1$Instituto de Ciencias
  Nucleares, Universidad Nacional Aut\'onoma de M\'exico, Apartado
  Postal 70-543, M\'exico Distrito Federal 04510,
  Mexico.\\
  $^2$Centre for Theoretical and Mathematical Physics, and Department of Physics,
  University of Cape Town, Rondebosch 7700, South Africa\\
  $^3$Instituto de F\1sica, Pontificia Universidad Cat\'olica de Chile,
  Casilla 306, Santiago 22, Chile.\\
  $^4$Centro Cient\1fico-Tecnol\'ogico de Valpara\1so, Casilla 110-V, Valpara\1so, Chile.\\
  $^5$Centro de Investigaci\'on y Desarrollo en Ciencias Aeroespaciales (CIDCA), Fuerza A\'erea de Chile, Santiago, Chile. \\
  $^6$Instituto de Ciencias B\'asicas, Universidad Diego Portales, Casilla 298-V, Santiago, Chile}

\begin{abstract}

We compute the vacuum one-loop quark-gluon vertex correction at zero temperature in the presence of a magnetic field. From the vertex function we extract the effective quark-gluon coupling and show that it grows with increasing magnetic field strength. The effect is due to a subtle competition between the color charge associated to gluons and the color charge associated to quarks, the former being larger than the latter. In contrast, at high temperature the effective thermo-magnetic coupling results exclusively from the contribution of the color charge associated to quarks. This produces a decrease of the coupling with increasing field strength. We interpret the results in terms of a geometrical effect whereby the magnetic field induces, on average, a closer distance between the (electrically charged) quarks and antiquarks. At high temperature, since the effective coupling is proportional only to the color charge associated to quarks, such proximity with increasing field strength makes the effective coupling decrease due to asymptotic freedom. In turn, this leads to a decreasing quark condensate. In contrast, at zero temperature both the effective strong coupling and the quark condensate increase with increasing magnetic field. This is due to the color charge associated to gluons dominating over that associated to quarks, with both having the opposite sign. Thus, the gluons induce a kind of  screening of the quark color charge, in spite of the quark-antiquark proximity. We discuss the implications for the inverse magnetic catalysis phenomenon.

\end{abstract}


\end{frontmatter}



The properties of strongly interacting matter in the presence of magnetic fields, as found in recent lattice QCD (LQCD) determinations, exhibit intriguing characteristics. In a thermal environment, at and above the transition temperature for deconfinement/chiral symmetry restoration, the magnetic field hinders the formation of the quark condensate~\cite{Bali} and makes the critical temperature  decrease with increasing field strength~\cite{Fodor}. This behavior is dubbed {\it inverse magnetic catalysis}. In contrast, the vacuum ($T=0$) condensate grows with the magnetic field strength. As the temperature increases near, but below the transition temperature, the condensate begins to grow  for weak fields reaching a maximum value, smaller than for $T=0$ and the same field strength. Subsequently, the condensate decreases with increasing field strength. This growth of the quark condensate with magnetic field strength corresponds to magnetic catalysis. Overall, this behavior indicates that the strength of the QCD interaction at $T=0$ is enhanced by the magnetic field, thus strengthening the binding of quark-antiquark pairs that make up the condensate. However, as the temperature increases, such binding becomes weaker. When the temperature reaches the transition region the magnetic field dominates the interaction, quenching monotonically the binding for all field strengths. The search for an explanation of such properties has attracted the attention of a great deal of research over the last years~\cite{attention,decreasing}.  A possible way to look at this effect has been casted in terms of the competition between the valence and the sea contributions to the quark condensate. It has been argued that at $T = 0$ both contributions are growing as a function of $eB$.  However, around the critical temperature $T_c$ the valence contribution is still increasing whereas the sea contribution decreases, as a
function of $eB$. This seemingly results in a decrease of $T_c$ as  a  function of $eB$. For recent reviews see~\cite{review1,review2}.

On general grounds a magnetic field interacting with electrically charged particles acts as an {\it ordering agent.} In other words, the  motion of virtual or real charges takes place around the magnetic field lines. This ordered motion has an important geometrical consequence: charged particles are closer to each other on average. When the intensity of the magnetic field increases, so does the proximity between charges. As is well known, due to asymptotic freedom, the closer strongly interacting particles are, the weaker the interaction. However strongly interacting matter, either at zero or at finite temperature, is not only made out of quarks and antiquarks but also of electrically neutral gluons. If the geometrical effect produced by the magnetic field were related to inverse magnetic catalysis, then at low temperatures the color interactions produced by gluons should dominate, while quarks would take over at high temperatures.  

An important clue on the properties of strongly interacting matter in the presence of a magnetic field has been provided in~\cite{Ayala} for the case of high temperature. There it was shown that under such conditions the quark-gluon effective coupling decreases with the field intensity and that the color charge contribution from the gluons cancels exactly. Furthermore, the magnetic field-dependent vertex correction satisfies a Ward-like identity involving the magnetic field dependent quark self-energy. This means that at high temperature color dynamics is dominated by quarks. This behavior can be understood in terms of the geometrical picture whereby the proximity between electric charges induced by the magnetic field dominates the color interaction. An outstanding question is whether this picture holds also at $T=0$, namely, whether under such circumstances the strength of the color interaction becomes, instead, gluon dominated.

In this paper we compute the magnetic field contribution to the quark-gluon vertex in vacuum and show that, indeed, the strong interaction becomes dominated by the contribution of the electrically neutral gluons. This generates an effective coupling that grows with increasing field strength, in contrast with the high-temperature result. Recall that inverse magnetic catalysis can also be quantified in terms of the properties of the quark condensate as a function of the magnetic field. Since the condensate is a measure of the strength of the bound between either vacuum $(T=0)$ or thermal $(T\neq 0)$ quark-antiquark pairs and $\alpha_s$ is a measure of the strength of the interaction between these quark-antiquark pairs, both quantities represent the strength of the quark-antiquark binding. We show that a mechanism that can help understand inverse magnetic catalysis consists on pursuing the relation between the properties of $\alpha_s$ as a function of the magnetic field and the condensate. In this context we recall that several calculations that address the behavior of the quark condensate in the presence of a magnetic field, coincide in that the condensate is an increasing function of the field strength~\cite{condincr}. Both, the coupling constant and the condensate, should behave similarly as a function of the field strength. We find that in the two extreme cases, namely, at high and zero $T$, they do. Here we do not address the details of how this change happens, which certainly require non-perturbative information for their description. However, by establishing that this change in the properties of $\alpha_s$ happens at these two extremes, we put forward a novel scenario to study inverse magnetic catalysis in terms of the thermomagnetic properties of the strong coupling constant.
 
We begin by considering the case of a magnetic field pointing along the $\hat{z}$ direction. In a magnetic background, the fermion propagator in coordinate space can no longer be written as a simple Fourier transform of a momentum propagator but instead it is written as~\cite{Schwinger}
\bea
   S(x,x')=\Phi (x,x')\int\frac{d^4p}{(2\pi)^4}e^{-ip\cdot (x-x')}S(p),
\label{genprop}
\eea
where $\Phi (x,x')$ is called the {\it Schwinger phase factor}. The translational invariant part of the propagator, $S(p)$, is given by
\begin{align}
   iS(p)&=\int_0^\infty \frac{ds}{\cos (qBs)} e^{is(p_{\|}^2-p_\perp^2
   \frac{\tan (qBs)}{qBs} - m^2)}\nn\\
   &\times \Big\{\left[\cos (qBs) + \gamma_1 \gamma_2 \sin (qBs) \right] (m+\slsh{p_{\|}}) \nn\\ 
   &- \frac{\slsh{p_\bot}}{\cos(qBs)} \Big\},
\label{Schwinger}
\end{align}
where $m$ and $q$ are the quark mass and absolute value of the quark charge, in units of the electron charge, respectively. Hereafter we use the following definitions for the parallel and perpendicular components of the scalar product of any two vectors $a^\mu$ and $b^\mu$
\begin{align}
   (a\cdot b)_{\|} &= a_0b_0 - a_3b_3\nn\\
   (a\cdot b)_{\bot} &= a_1b_1 + a_2b_2.
\label{defs}
\end{align}
%
%\begin{widetext}
%%%%%%%%%%%%%%%%%%%%%%%%%%%%%%%%%%%
\begin{figure*}[t]
\begin{center}
\includegraphics[scale=0.45]{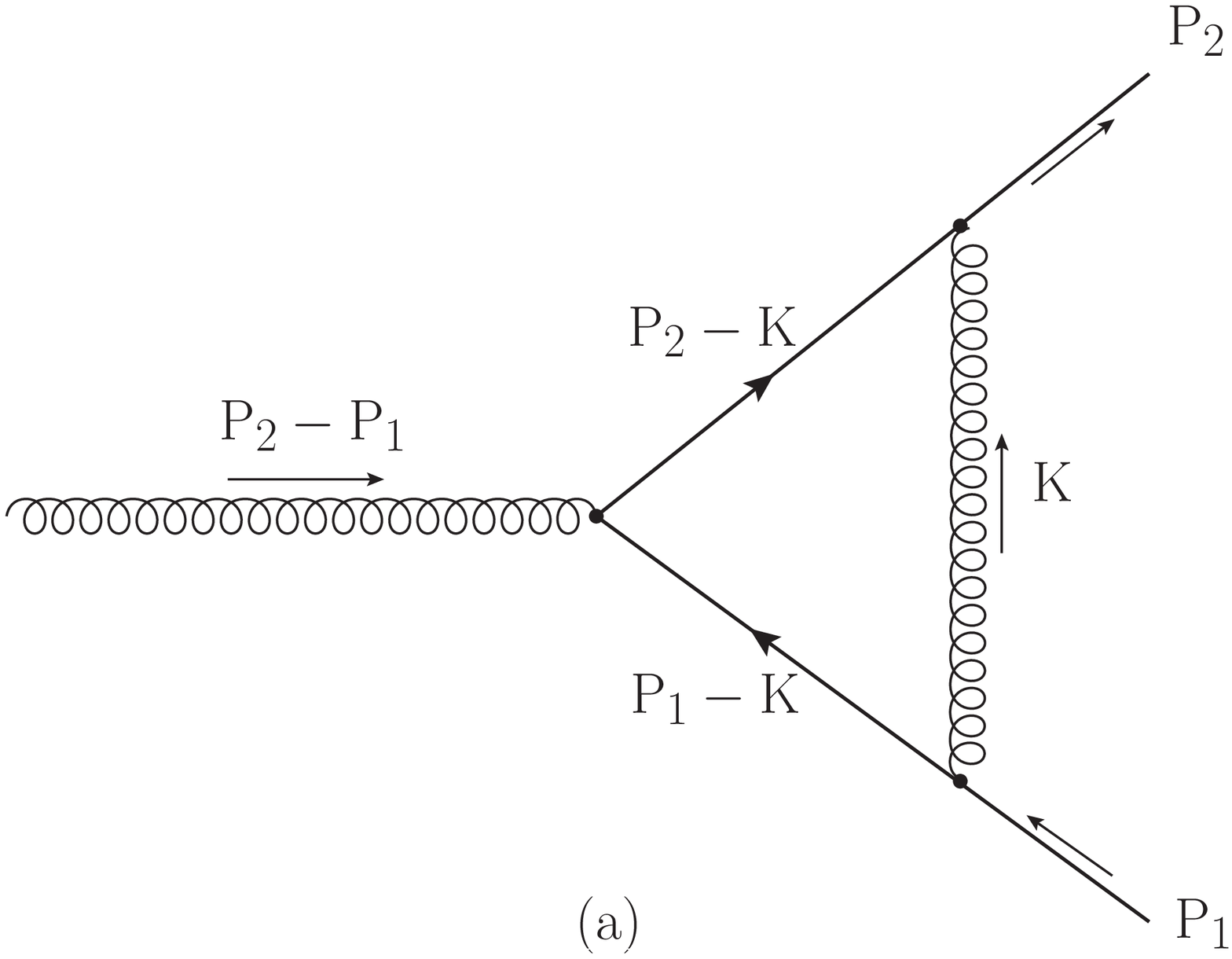}
\includegraphics[scale=0.45]{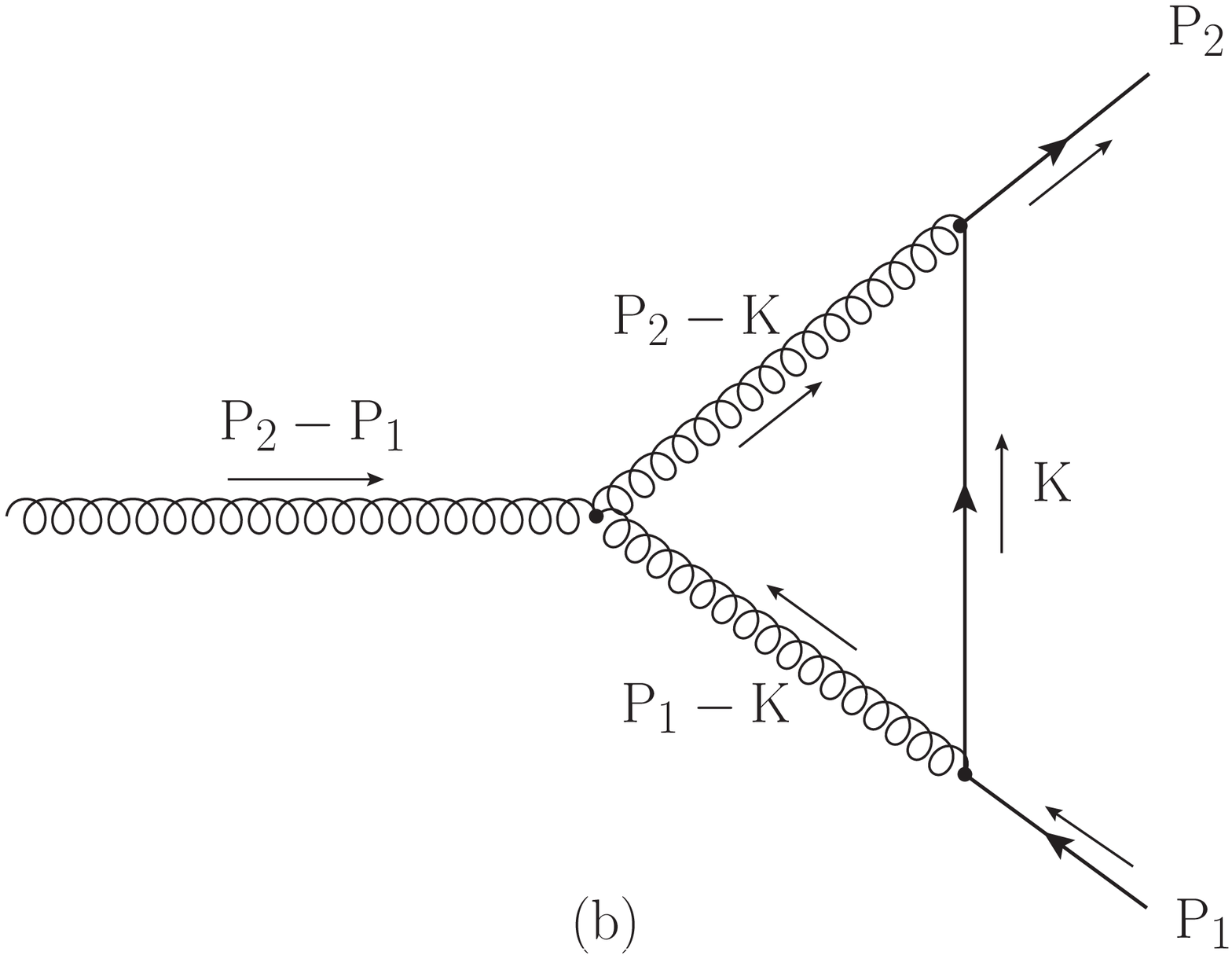}
\end{center}
\caption{Feynman diagrams contributing to the magnetic dependence of the quark-gluon vertex. Diagram (a) corresponds to a QED-like contribution whereas diagram (b) corresponds to a pure QCD contribution.}
\label{fig1}
\end{figure*}
%%%%%%%%%%%%%%%%%%%%%%%%%%%%%%%%%%%
%\end{widetext}

Figure~\ref{fig1} shows the Feynman diagrams contributing to the quark-gluon vertex. Diagram (a) corresponds to a QED-like contribution whereas diagram (b) corresponds to the pure QCD contribution. The computation of these diagrams requires the fermion propagator given by Eq.~(\ref{genprop}), which involves the Schwinger phase factor $\Phi (x,x')$. It can be shown~\cite{Ayala} that when only one or two fermion propagators are involved in this kind of triangle loop,
the phase factor can be {\it gauged away} and we can just work with the translationally invariant part of the fermion propagators.

Since the effect we are after shows up already for small magnetic field strengths, we  consider the case of a weak field for which the fermion propagator can be written as~\cite{Chyi}
\bea
   iS(p)=i\frac{\slsh{p}}{p^2} - (qB)\gamma_1\gamma_2\frac{\slsh{p}_\parallel}{p^4},
\label{weak}
\eea
where we consider the chiral limit, namely $m=0$. The chiral limit of the weak field expansion of the fermion propagator is a well defined object. In fact, this expansion can be viewed as a power series in $eB$ of the full propagator, independently of any relation between the field and the fermion mass. In the present context a  field is weak if compared with the gluon momentum squared, which must be large in a  perturbation calculation.

Working in the Feynman gauge, the contributions to the magnetic field dependent part of the quark-gluon vertex from diagrams $(a)$ and $(b)$ of Fig.~\ref{fig1}, in the weak field limit are
\begin{align}
   \delta\Gamma_{(a)}^{\mu\alpha}&= ig^3(qB)\left(C_F-\frac{C_A}{2}\right)t^\alpha
   \int\frac{d^4k}{(2\pi)^4}\frac{1}{k^2}\nn\\
   &\times \left\{
   \gamma^\nu\frac{(\slsh{p}_2-\slsh{k})}{(p_2-k)^2}\gamma^\mu
   \frac{\gamma_1\gamma_2(\slsh{p}_1-\slsh{k})_\parallel}{(p_1-k)^4}\gamma_\nu
   \right.\nn\\
   &+ \left. \gamma^\nu\frac{\gamma_1\gamma_2(\slsh{p}_2-\slsh{k})_\parallel}{(p_2-k)^4}
   \gamma^\mu\frac{(\slsh{p}_1-\slsh{k})}{(p_1-k)^2}\gamma_\nu\right\},
   \label{Gammaa-b1}
\end{align}   
   
\begin{align}   
   \delta\Gamma_{(b)}^{\mu\alpha}&=-2ig^3(qB)\frac{C_A}{2}t^\alpha
   \int\frac{d^4k}{(2\pi)^4}\frac{1}{k^4}\left[ g^{\mu\nu}(2p_2-p_1-k)^\rho \right.\nn\\
   &+\left. g^{\nu\rho}(2k-p_2-p_1)^\mu + g^{\rho\mu}(2p_1-k-p_2)^\nu \right]\nn\\
   &\times\gamma_\rho\frac{\gamma_1\gamma_2\slsh{k}_\parallel}
   {(p_2-k)^2(p_1-k)^2}\gamma_\nu,
\label{Gammaa-b2}
\end{align}
where $C_F$, $C_A$ are the color factors corresponding to the fundamental and adjoint representations of the $SU(N)$ Casimir operators, $C_F=(N^2-1)/2N$, and $C_A=N$ and $t^\alpha$ is a Gell-Mann matrix. The explicit factor of 2 in  Eq.~(\ref{Gammaa-b2}) takes care of the two possible charge fluxes in diagram $(b)$ of Fig.~\ref{fig1}.

We consider $\Gamma_{(a)}^{\mu\alpha}$ and $\Gamma_{(b)}^{\mu\alpha}$ as functions of the relative and average quark-pair four-momenta, $Q=p_1-p_2$ and $P=(p_1+p_2)/2$, respectively. According to the kinematics depicted in Fig.~\ref{fig1}, $Q$ corresponds to the four-momentum carried by the gluon. For simplicity we consider the symmetric three-momentum configuration where $p_1=(E,\vec{p})$, $-p_2=(E,-\vec{p})$, thus $Q=(2E,\vec{0})$ and $P=(0,\vec{p})$. In this case, $Q^2$ is proportional to the energy and $P^2$ to the momentum squared carried by the gluon. To make a closer connection to the case discussed in Ref.~\cite{Ayala}, we work in the {\it static} limit, namely $P\to 0$. Furthermore, in order to make sure that the perturbative calculation makes sense, we take $Q^2$ large. In this sense, the expansion parameter for the validity of the calculation becomes $qB/Q^2$. In this limit, after a lengthy but straightforward exercise,  Eqs.~(\ref{Gammaa-b1})-(\ref{Gammaa-b2}) become
\begin{align}
   \delta\Gamma_{(a)}^{\mu\alpha}&=-ig^3\left(C_F-\frac{C_A}{2}\right)t^\alpha
   \frac{\left[1+\ln(4)\right]}{3\pi^2}\nn\\
   &\times\frac{q\vec{\Sigma}\cdot\vec{B}}{Q^2}\left(\slsh{u}u^\mu + \slsh{b}b^\mu\right)\;,
   \label{Gammaa-b-become1}
\end{align}   
\begin{align}   
   \delta\Gamma_{(b)}^{\mu\alpha}&=-ig^3C_At^\alpha
   \frac{\left[-1+\ln(4)\right]}{15\pi^2}\nn\\
   &\times\frac{q\vec{\Sigma}\cdot\vec{B}}{Q^2}\left(\slsh{u}u^\mu + \slsh{b}b^\mu\right),
\label{Gammaa-b-become2}
\end{align}
where $\vec{\Sigma}\cdot\vec{B}=\Sigma_3B=i\gamma_1\gamma_2B$ is the dot product between the spin operator and the magnetic field vector and we have defined $u^\mu=(1,0,0,0)$ and $b^\mu=(0,0,0,1)$. Notice that the first order magnetic field-dependent correction is proportional to the coupling between the quark spin and the magnetic field, affecting only the longitudinal components $(\mu=0,3)$. The same longitudinal matrix structure has been found for the vertex correction in the presence of a magnetic field in the context of an effective QCD model~\cite{Ferrer1} and in QED~\cite{Ferrer2}.

From the longitudinal components of the full vertex (to this order), namely 
\bea
   \Gamma_\parallel^\alpha=i\gamma^\mu_\parallel t^\alpha + \delta\Gamma_{(a)}^{\mu\alpha} 
   + \delta\Gamma_{(b)}^{\mu\alpha},
\label{fullong}
\eea
one can extract the effective vacuum quark-gluon coupling in the presence of a magnetic field
\begin{align}
   g_{\mbox{\small{eff}}}^{\mbox{\small{vac}}}&=g-\left[g^3\frac{1}{3\pi^2}\frac{q\vec{\Sigma}\cdot\vec{B}}
   {Q^2}\right]\nn\\
   &\times
   \left\{ \left(C_F-\frac{C_A}{2}\right)\left[1+\ln(4)\right] + \frac{C_A}{5}\left[-1+\ln(4)\right]\right\}\nn\\
   &=g-\left[g^3\frac{1}{3\pi^2}\frac{q\vec{\Sigma}\cdot\vec{B}}{Q^2}\right]\nn\\
   &\times
   \left\{ \left[1+\ln(4)\right]C_F - \frac{\left[7+3\ln(4)\right]}{10}C_A\right\}.
\label{geff}
\end{align}
For $N=3$, the contribution from the color charge associated to gluons $(C_A)$ dominates over the contribution from the color charge associated to quarks $(C_F)$. The net effect is that in vacuum, the effective coupling between quarks and gluons grows with the magnetic field strength. In contrast, we recall that the effective thermo-magnetic coupling computed at high temperature becomes~\cite{Ayala}
\begin{align}
   g_{\mbox{\small{eff}}}^{\mbox{\small{therm}}}=g\left[1-\frac{m_f^2}{T^2}+\left(\frac{8}
   {3T^2}\right)g^2C_FM^2(T,m_f,qB)\right],\nn\\
\label{effective}
\end{align}
where $m_f$ is the quark thermal mass and the function $M^2(T,m,qB)$ is given by
\bea
   M^2(T,m,qB)=\frac{q\vec{\Sigma}\cdot\vec{B}}{16\pi^2}\left[\ln(2) - \frac{\pi}{2}\frac{T}{m}\right],
\label{M}
\eea
which for high temperature is negative definite. Notice that contrary to the $T=0$ case, the magnetic field-dependent correction at high temperature is proportional only to the contribution from the color charge associated to quarks, {\it i.e.} $C_F$. This is because the contribution from the color charge associated to gluons, $C_A$, cancels identically. 

Equations~(\ref{geff}) and~(\ref{effective}) show that in the presence of a magnetic field, at $T=0$, the contribution from the color charges associated to gluons  dominates marginally over the contribution from the color charge associated to quarks.  Since the former has the opposite sign of the latter, the overall effective coupling grows with the magnetic field strength. At high temperature however, the contribution from the color charge associated to gluons cancels and the color dynamics is quark-dominated. Since the surviving magnetic field-dependent contribution has an overall  negative sign, the effective coupling decreases with the magnetic field strength. We point out that calculations carried out in the opposite limit, namely the very strong field case, find that the coupling constant at $T=0$ decreases as a function of the field strength~\cite{Miransky-alpha}. Altogether this means that the behavior found in this work should be valid up to a certain (albeit large) value of the magnetic field.

Notice that the perturbative calculation at $T=0$ requires that $Q^2$ is large and that the weak field approximation is valid provided $qB\ll Q^2$. At finite temperature, the large temperature assumption provides the large energy scale for the perturbative calculation (Hard Thermal Loop approximation) as well as for the weak field approximation to be valid. 

Also, notice that the kinematical conditions we have implemented include studying the configuration where the quark and antiquark travel back to back. This means that their relative orbital angular momentum $L$ vanishes. Since the gluon spin is $S=1$, the quarks must carry a total spin $S=1$ with a preferred projection aligned with the magnetic field direction. Had we considered a different kinematical configuration whereby the quark-antiquark pair emerged with another relative angle different from 180 degrees, conservation of angular momentum and parity implies that the relative angular momentum $L$ has to be either 0 or 2. In both cases, the total quark-antiquark spin needs to be $S=1$. 

Also, we point out that our calculation provides not only the behavior of the effective coupling constant but also of the effective quark-gluon vertex as a function of the magnetic field (in the weak field limit). This vertex can in turn be used to compute a given process that may be influenced by the presence of the magnetic field. Consider for instance $\bar{q} q \rightarrow \bar{q} q$. Using the effective vertex found in this work, the amplitude for this process can be constructed attaching the gluon line to the incoming $\bar{q} q$ whereas the outgoing $\bar{q} q$ is already provided by the vertex. The process can be described in any given Lorentz frame. We thus see that choosing the symmetric configuration is tantamount to working in the center of mass of the colliding pair. Since the matrix element is Lorentz invariant, the choice of frame is a matter of convenience. The use of the static limit is an approximation that is valid provided there is a large scale (larger than the quark momenta or the masses) present in the calculation. This large scale is the gluon virtuality $Q^2$. When this quantity is large so it is the energy of the collision in the above-described process. This means that the calculation lends itself to be applied to describing hard $\bar{q} q$ annihilation (or scattering). This kind of processes are relevant in collisions of hadronic systems, namely A+p or p+p and even A+A with a large momentum transfer involved, where the energy is larger than the temperature, if any. In summary, the choice of configuration and of kinematics is general enough under these circumstances. 

Finally, notice that the study is performed by looking at two extreme scenarios where perturbation theory at leading order is under control, therefore avoiding the ambiguities of non-perturbative elements where modeling is oftentimes involved (see for example Ref.~\cite{suggested}).  In these limits a first order calculation in the magnetic field intensity suffices for two reasons: First, since there is a large energy scale provided either by the temperature (squared) or by the quark's momentum (squared), the field can be taken as small with respect to either of these energy scales. Second, the LQCD calculation for the condensate in the (high) zero temperature limit is a monotonically (decreasing) increasing function of the field strength. In order to study if $\alpha_s$ behaves similarly with the magnetic field strength, what matters is knowledge of the sign of the first derivative of $\alpha_s$ at $qB=0$. This can be computed merely from the linear term in $qB$ which is the term computed in this work. In summary, although interesting effects take place in the opposite limit, namely the strong field case (see for example Ref.~\cite{Endrodi}), for the purpose of this work, as argued, it suffices to work in the weak field limit. In the same context, applying a standard renormalization group analysis to explore the change of the coupling with scale will not affect the sign of its rate of change with the magnetic field.

Our results show that the geometrical effect produced by the magnetic field at high temperature, whereby  quarks and anti-quarks get closer on average, is accompanied by the decrease of their effective interaction due to asymptotic freedom. This  takes place because in that scenario the strong interactions are due entirely to the color charge associated to quarks. The strength of the interaction thus decreases with increasing magnetic field strength. In contrast, at $T=0$ such geometrical effect does not take place. This is because the color charge associated to gluons produces a kind of screening of the color charge associated to quarks. In turn, and in spite of the quark-anti-quark proximity, this leads to an increase in the effective strong coupling with  increasing magnetic field strength. Such larger coupling  results in a tighter quark-anti-quark bond, leading to a larger quark condensate as obtained in LQCD at $T=0$. In contrast, a smaller coupling translates into a looser quark-anti-quark bond and thus into a decreasing condensate at large $T$, as also found by LQCD.  Similar considerations phrased in terms of the competition between valence and sea-quark contributions around $T_c$ have been argued in Ref.~\cite{review1}.

The details of how this change in behavior of the coupling constant take place in the intermediate (non-perturbative domain) with increasing field strength as well as its relation to the behavior of the critical temperature above and below $T_c$ are still open problems. Work along these directions is in progress and will be reported elsewhere.

\section*{Acknowledgments}

This work has been supported in part  by  CONACyT (M\'exico)  grant number 128534,  UNAM-DGAPA-PAPIIT grant number IN101515,   National Research Foundation (South Africa), the Harry Oppenheimer Memorial Trust OMT Ref. 20242/02, and  by FONDECYT (Chile) grant numbers 130056, 1120770, and 1150847. LAH acknowledges the University of Cape Town for funding assistance.





\end{document}